\documentclass[journal,comsoc]{IEEEtran}

\usepackage{graphicx}
\usepackage{epstopdf}
\usepackage{float}
\usepackage{algorithm,algorithmic}
\usepackage{array}
\usepackage{amsmath}
\usepackage{amssymb}
\usepackage{mdwmath}
\usepackage{mdwtab}
\usepackage{eqparbox}

\usepackage{fixltx2e}
\usepackage{cases}
\usepackage{bm}
\usepackage{multirow}
\usepackage{psfrag}
\usepackage[usenames]{color}
\usepackage{mathtools}

\usepackage{fixmath}
\usepackage{mathdots}
\usepackage{latexsym}
\usepackage{amsmath,amssymb}

\usepackage{bm}
\usepackage{footmisc}

\usepackage[T1]{fontenc}
\usepackage{url}
\usepackage{cite}
\newcommand{\subparagraph}{}
\usepackage{titlesec}

\newtheorem{theorem}{\bf Theorem}

\newcommand{\diag}{\mathop{\textrm{diag}}}
\newcommand{\tr}{\mathop{\textrm{Tr}}}

\begin{document}
\title{MIMO Capacity with Reduced RF Chains}
\author{Shuaishuai~Guo,~\IEEEmembership{Member, IEEE,}
Haixia~Zhang,~\IEEEmembership{Senior Member, IEEE,}\\ and   Mohamed-Slim Alouini,~\IEEEmembership{Fellow, IEEE}
\thanks{S. Guo and M. -S. Alouini are with  King
Abdullah University of Science and Technology (KAUST), Thuwal, Makkah
Province, Kingdom of Saudi Arabia, 23955-6900 (email: shuaishuai.guo@kaust.edu.sa;  slim.alouini@kaust.edu.sa).}
\thanks{H. Zhang is with Shandong Provincial Key Laboratory of Wireless Communication Technologies, Shandong University, Jinan, China, 250061 (e-mail: haixia.zhang@sdu.edu.cn).}
\thanks{Partial of the work has been submitted to ITA Workshop 2019 for presentation \cite{Guo2019a}.}}
\markboth{Submitted to IEEE TRANSACTIONS ON COMMUNICATIONS}%
{Guo \MakeLowercase{\textit{et al.}}:MIMO Capacity with Reduced RF Chains}
\maketitle

\begin{abstract} 
This paper presents a remarkable advance for the understanding of MIMO capacity limits with insufficient RF chains. The capacity is characterized by the maximum mutual information given any vector inputs subject to not only an average power constraint but also a sparsity constraint.  It is proven that the Gaussian mixture input distribution is capacity-achieving in the high signal-to-noise-ratio (SNR) regime. The optimal mixture coefficients and the covariance matrices of the Gaussian mixtures are derived and also the corresponding achievable capacity. For the special case with a single RF chain, the optimal mixture coefficients are shown to be approximately proportional to channel gains. The capacity is approximately the maximum-ratio combining (MRC) of multiple channels in the high SNR regime.
 We investigate the superiority of capacity-achieving techniques: Non-Uniform Spatial Modulation (NUSM) and Non-Uniform Beamspace Modulation (NUBM) by comparing them with the best antenna/beamspace selection (BAS/BBS) and the uniform spatial/beamspace modulation (USM/UBM). The comparison results reveal that the information-guided NUSM/NUBM is optimal in the high SNR regime.  Numerical results are presented to validate our analysis.

\end{abstract}


\IEEEpeerreviewmaketitle

\section{Introduction}
\IEEEPARstart{C}{hannel} capacity characterizes the maximum error-free throughput of a channel
given any signal input subject to some specific constraints. From both theoretical and practical standpoints, channel capacity analysis is with extreme importance, since it can give many insights into the information-carrying capabilities of a channel and can offer significant design guidelines for communication systems. However, only a few  of channels are known on capacity because finding the constrained capacity-achieving input distribution is a rather intricate task. The famous Shannon capacity for a single-input single-output (SISO) channel was derived on the premise of an average power-constrained input in 1940-1950s \cite{Shannon1948,Shannon1949}. The zero-mean complex Gaussian distribution is obtained as the optimal input distribution and the channel capacity of a SISO channel is given by
\begin{equation}
\mathsf{C}_{\textrm{SISO}}=B\log_2 \left(1+\gamma\right),
\end{equation}
where $B$ denotes the spectrum bandwidth in hertz and $\gamma$ stands for the receive signal-to-noise-ratio (SNR). Shannon capacity is one of the cornerstones of modern communication networks and also a theoretical basis in the derivation of other channel capacities. For instance, the multiple-input multiple-out (MIMO) channel capacity with perfect channel state information at the transmitter (CSIT) was derived based on the SISO capacity as well as the channel 
parallelization \cite{Telatar1999,Goldsmith2003}.

Knowing the capability of MIMO structures in improving channel throughput has facilitated a wide investigation from both academia and industry during the past decade. Now, the antenna scales  at the transceiver sides are growing larger and larger for multiplexing, diversity, beamforming gains, etc \cite{Marzett2010,Larsson2014}. However, a concern has emerged on the implementation cost and power consumption of the associated massive number of radio frequency (RF) chains, especially for the communications at high frequencies with broad bandwidth such as the promising millimeter wave (mmWave) communications and  Terahertz communications that enable the next generation of mobile communication networks.
In MIMO systems with reduced RF chains, there are two  types of connections between the RF chains and the transmit antennas:  antenna switches or  beamformers (i.e., precoders) as illustrated in Figs. 1(a)-(b). In such MIMO systems, redundant antennas/beampaces exist, but the number of RF chains limits the number of data streams. 

\begin{figure}[t]
  \centering
  \includegraphics[width=0.42\textwidth]{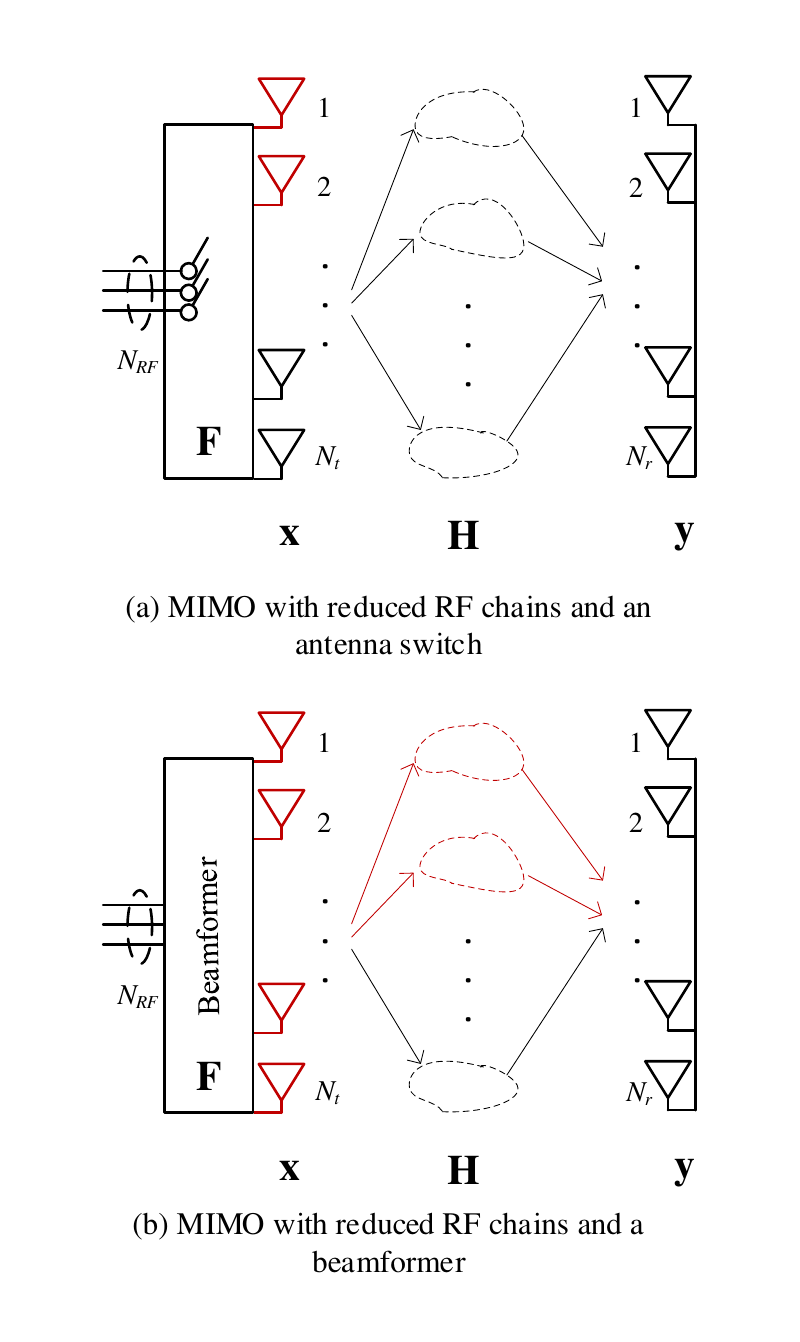}\\
 \caption{MIMO systems with insufficient RF chains employing either an antenna switch or a beamformer as the connector between RF chains and transmit antennas.}
  \label{System_Model}
\end{figure}

In MIMO systems associated with insufficient RF chains, the signal input  $\mathbf{x}$  to the channel can be written as $\mathbf{x}=\mathbf{F}\mathbf{s}$, where $\mathbf{F}$ is the antenna selection indicator matrix or the beamformer in Figs. \ref{System_Model} (a)-(b) and $\mathbf{s}$ is the input to RF chains.
Previous understanding of capacity only took the average power constraint on $\mathbf{s}$ into consideration, which have motivated researchers to select the best antenna/beamspace combination to convey information  \cite{Molisch2005,Ayach2014,Dai2015,Li2017,Molu2018,Yu2016,He2018}. Mathematically, the understanding can be  expressed by
\begin{equation}\label{PreCap}
\mathsf{C}_{\mathbf{H}}=\max_{f_{\mathbf{x}}(\mathbf{x})}\mathsf{I}(\mathbf{x};\mathbf{y}|\mathbf{H})=\max_{\mathbf{F}}\max_{\mathbf{f}_{\mathsf{S}}(\mathsf{s})}\mathsf{I}(\mathbf{s};\mathbf{y}|\mathbf{H}),
\end{equation}
where $\mathbf{H}$ is a constant channel matrix and $\mathbf{y}$ is the output of the channel.
Based on the understanding of capacity limit,  the best $\mathbf{F}$ is used for transmission, known as best antenna selection (BAS) or best beamspace selection (BBS) for two connection structures, respectively. BAS adopts the best transmit antenna combination (TAC) while BBS utilizes the best singular vector space. Both of them were widely recognized as upper bounds of the achievable rate in \cite{Molisch2005,Ayach2014,Dai2015,Li2017,Molu2018,Yu2016,He2018}, until they have been recently challenged by a branch of MIMO techniques: spatial modulation (SM) \cite{Mesleh2008,Yang2008,Renzo2014,Guo2016,Guo2017,Ibrahim2016,Basnayaka2016,Faregh2016,Narasimhan2016,Younis2017,Younis2018} in systems using antenna switches as connectors and beamspace modulation (BM) \cite{Guo2019} in systems using beamformers as connectors.
To analyze the capacity, the authors of \cite{Younis2017} and \cite{Younis2018} have recently proposed to use the joint distribution $f_\mathbf{HFs}(\mathbf{HFs})$ to convey information and derived the product $\mathbf{H}\mathbf{F}\mathbf{s}$ to be Gaussian distributed for achieving the capacity. Even though a higher system throughout was obtained regardless of any channel distribution and any number of transmit antennas, it should be mentioned that $\mathbf{H}$ should not be viewed as the input to the channel and also the receiver typically knows the instantaneous $\mathbf{H}$ for information decoding.   These conflicts motivate us to rethink the channel capacity limits of  MIMO systems with reduced RF chains. The contributions of our work are listed as follows.

\begin{itemize}
\item The MIMO capacity with insufficient RF chains is characterized by the maximum mutual information given any vector inputs subject to not only an average power constraint but also a sparsity constraint.

\item  It is proven that the Gaussian mixture input distribution is capacity-achieving in the high SNR regime when the receive degrees of freedom (RDOFs) are larger than the number of RF chains. We obtain the optimal mixture coefficients and the covariance matrices of the Gaussian mixtures as well as the capacity expression in the high SNR regime.

\item For the special case with a single RF chain, the optimal mixture coefficients are shown to be approximately proportional to channel gains. We reveal that the capacity is approximately the maximum-ratio combining (MRC) of multiple channels in the high SNR regime.
\item We investigate the superiority of capacity-achieving techniques: Non-Uniform Spatial Modulation (NUSM) \cite{Liu2016} and Non-Uniform Beamspace Modulation \cite{Guo2019} by comparing them with BAS/BBS and uniform spatial/beamspace modulation (USM/UBM) \cite{Mesleh2008,Yang2008,Renzo2014,Guo2016,Guo2017,Ibrahim2016,Basnayaka2016,Faregh2016,Narasimhan2016}. 
Numerical results are presented to validate our analysis.
\end{itemize}

The rest of the paper is organized as follows. The system model is depicted in Section II. In Section III, we discuss the  constraints on the input vectors to the equivalent channels and deduce the capacity-achieving input distribution  as well as the expression of capacity in the high SNR regime. To give insights, examples are given in Section IV to show the improvement of the new understanding. In this section, we validate the superiority of NUSM and NUBM. In addition, we also show conventional USM/UBM outperforms BAS/BBS only under some specific channel conditions in this section, while NUSM/NUBM is certain to outperform BAS/BBS and USM/UBM. 
We conclude the whole paper in the last section.

In this paper, we use $x$, $\mathbf{x}$, $\mathbf{x}$, $\mathcal{X}$ to denote a scaler, a vector, a matrix and a set, respectively. $||\mathbf{x}||_0$ and $||\mathbf{x}||_2$ are adopted to represent $l_0$ and $l_2$ norms of $\mathbf{x}$, respectively. $\mathrm{dim}(\mathbf{x})$ is the dimension of $\mathbf{x}$. $(\cdot)^H$ denote the conjugate transpose. $\det(\mathbf{X})$ means the determination of $\mathbf{X}$. $|\mathcal{X}|$ stands for the size of $\mathcal{X}$.  $\mathsf{H}(\cdot)$ represents the entropy function. $\mathsf{I}(\mathbf{x};\mathbf{y})$ means the entropy between $\mathbf{x}$ and $\mathbf{y}$. $\mathbf{I}_N$ denotes an $N\times N$ identity matrix. $\mathcal{CN}(\mu,\sigma^2)$ represents complex Gaussian distribution with zero mean and variance $\sigma^2$. $\mathbb{E}(\cdot)$ is the expectation operation. $f_{\mathbf{x}}(\mathbf{x})$ stands for the probability density function (PDF) of $\mathbf{x}$. $(x)^+$ equals to $\max\{x,0\}$. 

\section{System Model}

Over point-to-point MIMO channels, we consider general $(N_t,N_R,N_{RF})$ MIMO systems with either antenna switches or beamformers as the connectors between RF chains and transmit antennas as depicted in Fig. \ref{System_Model} (a)-(b), where $N_t$, $N_r$ $N_{RF}$ represents the number of transmit antennas, the number of receive antennas and the number of RF chains, respectively.  It is assumed that $N_{RF}<N_t$ and $N_{RF}<N_r$; the signals are transmitted over narrowband flat fading channels; the channel state information is perfectly known at the transmitter and the receiver. For convenience, we assume that the bandwidth $B=1$ Hz and dedicate our work in analyzing the limit on spectral efficiency. Let $\mathbf{H}\in\mathbb{C}^{N_r\times N_t}$ denote channel matrix,
and the receive signal vector $\mathbf{y}\in\mathbb{C}^{N_{r}}$ can be expressed as
\begin{equation}
\mathbf{y}=\gamma\mathbf{H}\mathbf{x}+\mathbf{n}=\gamma\mathbf{H}\mathbf{Fs}+\mathbf{n},
\end{equation}
where $\gamma$ denotes the receive SNR: $\mathbf{x}=\mathbf{F}\mathbf{s}\in\mathbb{C}^{N_t}$ represents the input vector to the MIMO channels, which is assumed to be subject to a unit average power constraint $\tr(\mathbf{x}\mathbf{x}^H)\leq1$; $\mathbf{F}^{N_t\times N_{RF}}$ stands for the antenna selection indicator or the beamformer in Figs. \ref{System_Model} (a)-(b); $\mathbf{s}\in\mathbb{C}^{N_{RF}}$ is the input to RF chains; $\mathbf{n}\in\mathbb{C}^{N_r}$ denotes the additive Gaussian noise with zero mean and variance $\mathbf{I}_{N_r}$, i.e., $\mathbf{n}\sim \mathcal{CN}(\mathbf{0},\mathbf{I}_{N_r})$.

\section{Capacity Limits and Capacity-Achieving Distribution}

\subsection{Problem Formulation}
In MIMO systems with antenna switches as connectors, it is obvious that the transmitted signal vectors $\{\mathbf{x}\}$  are power-constrained and sparsity-constrained, because the number of activated antennas should be less than or equal to the number of RF chains. That is,
\begin{equation}
\tr\left(\mathbf{x}\mathbf{x}^H\right)\leq1
\end{equation}
and 
\begin{equation}
||\mathbf{x}||_0\leq N_{RF}.
\end{equation}
In MIMO with beamformers as connectors, based on the channel parallelization \cite{Telatar1999}, the capacity can be expressed as
\begin{equation}\label{RForm}
\begin{split}
\mathsf{C}_{\mathbf{H}}&=\max_{f_{\mathbf{x}}(\mathbf{x})}\mathsf{I}(\mathbf{x};\mathbf{y}|\mathbf{H})\\&=\max_{f_{\mathbf{X'}}(\mathbf{x'})}\mathsf{I}(\mathbf{V}\mathbf{x'};\mathbf{U}^H\mathbf{y}|\mathbf{H})
\\&=\max_{f_{\mathbf{X'}}(\mathbf{x'})}\mathsf{I}(\mathbf{x'};\mathbf{y'}|\mathbf{\Lambda}),
\end{split}
\end{equation}
where $\mathbf{x}=\mathbf{V}\mathbf{x'}$; $\mathbf{y'}=\mathbf{U}^H\mathbf{y}=\mathbf{\Lambda}\mathbf{x'}+\mathbf{U}^H\mathbf{n}$; $\mathbf{U}\in\mathbb{C}^{N_r\times m}$ and $\mathbf{V}\in\mathbb{C}^{N_t\times m}$ are left-singular and right-singular matrices of $\mathbf{H}$; $m=\textrm{rank}(\mathbf{H})$; $\mathbf{\Lambda}\in\mathbb{C}^{m\times m}$ is a diagonal matrix composed of the non-zeros singular values of  $\mathbf{H}$, that is,
\begin{equation}
\mathbf{H}=\mathbf{U}\mathbf{\Lambda}\mathbf{V}^H.
\end{equation}
Even though the real input vector $\mathbf{x}$ to the MIMO channel $\mathbf{H}$  is only power-constrained and not sparsity-constrained,   the input vector $\mathbf{x'}$ to the equivalent MIMO channel $\mathbf{\Lambda}$ is both power-constrained and sparsity-constrained, which can be given by
\begin{equation}
\tr\left(\mathbf{x'}\mathbf{x'}^H\right)\leq1,
\end{equation}
and 
\begin{equation}
||\mathbf{x'}||_0\leq N_{RF}.
\end{equation}

Therefore, summarizing all above,  we conclude that for MIMO systems associated with insufficient RF chains employing either antenna switches or beamformers as connectors, one can solve a  mutual information maximization problem given any vector inputs subject to  not only an average power constraint and a sparsity constraint to deduce the capacity. 

Since $\mathbf{x}$ and $\mathbf{x'}$ are sparse, they can expressed as 
\begin{equation}\label{eqAS}
\mathbf{x}=\mathbf{E}_i\mathbf{s},
\end{equation}
and 
\begin{equation}
\mathbf{x'}=\mathbf{E}_i'\mathbf{s},
\end{equation}
respectively, where $\mathbf{E}_i\in\mathbb{C}^{N_t\times N_{RF}}$ and $\mathbf{E}_i'\in\mathbb{C}^{m\times N_{RF}}$ play the role of non-zero position indicator; each column of  $\mathbf{E}_i$ is an $N_t$-dimensional vector basis with all zeros except an one and there are $N_c=\left(N_t\atop{N}_{RF}\right)$ feasible $\mathbf{E}_i$; 
 each column of  $\mathbf{E}_i'$ is an $m$-dimensional vector basis with all zeros except an one and there are $N_c'=\left(m\atop{N}_{RF}\right)$ feasible $\mathbf{E}_i'$.

In this paper, an important assumption is that the RDOFs are greater than the number of transmit RF chains. In systems employing antenna switches, it can be given by
\begin{equation}
\mathrm{dim}(\mathbf{y})=N_r>\mathrm{dim}(\mathbf{s})=N_{RF},
\end{equation}
and in systems employing beamformers, it is 
\begin{equation}
\mathrm{dim}(\mathbf{y'})=\mathrm{rank}(\mathbf{H})>\mathrm{dim}(\mathbf{s})=N_{RF}.
\end{equation}
The reason why this assumption is important is explained infra.
 Taking the systems connected with antenna switches as an example, the capacity of such systems can be expressed as
\begin{equation}
\begin{split}
\mathsf{C}_{\mathbf{H}}&=\max_{f_{\mathbf{x}}(\mathbf{x})}\mathsf{I}(\mathbf{x};\mathbf{y}|\mathbf{H})
\\&=\max_{f_{\mathbf{x}}(\mathbf{x})}\mathsf{H}(\mathbf{y})-\mathsf{H}(\mathbf{y}|\mathbf{H},\mathbf{x})
\\&=\max_{f_{\mathbf{x}}(\mathbf{x})}\mathsf{H}(\mathbf{y})-\mathsf{H}(\mathbf{n}),
\end{split}
\end{equation}
and maximizing the capacity is equivalent to maximizing the entropy of the received signal, i.e., $\mathsf{H}(\mathbf{y})$ \cite{Younis2017,Younis2018}. It is well known that $\mathbf{y}$ following a $\mathrm{dim}(\mathbf{y})$ independent complex Gaussian distribution  maximizes $\mathsf{H}(\mathbf{y})$ \cite{Cover2012}. In the case with $\mathrm{dim}(\mathbf{y})\leq \mathrm{dim}(\mathbf{s})$, because the dimension of data streams is sufficient to make  the information part $\mathbf{Hx}$ to follow a $\mathrm{dim}(\mathbf{y})$ independent complex Gaussian distribution and the capacity can be given by 
\begin{equation}
\mathsf{C}_{\mathbf{H}}=\max_{f_{\mathbf{x}}(\mathbf{x})}\mathsf{I}(\mathbf{x};\mathbf{y}|\mathbf{H})=\max_{\mathbf{E}_i}\max_{\mathbf{f}_{\mathbf{s}}(\mathsf{s})}\mathsf{I}(\mathbf{s};\mathbf{y}|\mathbf{H}),
\end{equation}
which is as same as (\ref{PreCap}). In the case with $\mathrm{dim}(\mathbf{y})>\mathrm{dim}(\mathbf{s})$, there are not sufficient independent data streams to make the information part $\mathbf{Hx}$ to follow a $\mathrm{dim}\{\mathbf{y}\}$ independent complex Gaussian distribution. Thus, it raises a question what are the optimal $f_{\mathbf{y}}(\mathbf{y})$ that maximizes $\mathsf{H}(\mathbf{y})$  and the corresponding constrained $f_{\mathbf{x}}(\mathbf{x})$ with the assumption $\mathrm{dim}(\mathbf{y})>\mathrm{dim}(\mathbf{s})$.

\subsection{Capacity-Achieving Input Distribution and Capacity Expression in the High SNR Regime}

As discussed in the last subsection, the capacity analysis for MIMO systems employing either antenna switches or beamformers as connectors is similar. Hence, we only take the capacity analysis for systems connected with antenna switches as an example to demonstrate the analysis.
Based on the expression of $\mathbf{x}$ in (\ref{eqAS}) and according the chain rule of mutual information \cite{Yang2008}, we can obtain
\begin{equation}
\begin{split}
\mathsf{I}(\mathbf{x};\mathbf{y}|\mathbf{H})&=\mathsf{I}(\mathbf{E}_i,\mathbf{s};\mathbf{y}|\mathbf{H})\\&=\mathsf{I}(\mathbf{s};\mathbf{y}|\mathbf{H})+\mathsf{I}(\mathbf{E}_i;\mathbf{y}|\mathbf{H},\mathbf{s})
\end{split}
\end{equation}
Therefore, the mutual information maximization problem  becomes
\begin{equation}
\begin{split}
\mathsf{C}_{\mathbf{H}}&=\max_{f_{\mathbf{x}}(\mathbf{x})}\mathsf{I}(\mathbf{x};\mathbf{y}|\mathbf{H})\\
&=\max_{f_{\mathbf{E}_i}({\mathbf{E}_i}),f_{\mathbf{s}|\mathbf{E}_i}(\mathbf{s}|\mathbf{E}_i)}\mathsf{I}(\mathbf{s};\mathbf{y}|\mathbf{H})+\mathsf{I}(\mathbf{E}_i;\mathbf{y}|\mathbf{H},\mathbf{s})
\end{split}
\end{equation}
Let $\mathcal{E}$ denote the candidate set of all feasible $\mathbf{E}_i$ with size $|\mathcal{E}|=N_c$. The discrete distribution of $\mathbf{E}_i$ can be represented by a probability distribution $\boldsymbol{\alpha}=[\alpha_1,\alpha_2,\cdots,\alpha_{N_c}]$, where
\begin{equation}
f_{\mathbf{E}_i}(\mathbf{E}_i)=P(\mathbf{E}_i)=\alpha_i,
\end{equation}
and
\begin{equation}
\sum_{\mathbf{E}_i\in\mathcal{E}}P(\mathbf{E}_i)=\sum_{i=1}^{N_c}\alpha_i=1.
\end{equation}
Then, the mutual information maximization problem becomes
\begin{equation}\label{eq18}
\begin{split}
\mathsf{C}_{\mathbf{H}}
&=\max_{\boldsymbol{\alpha},f_{\mathbf{s}|\mathbf{E}_i}(\mathbf{s}|\mathbf{E}_i)}\sum_{i=1}^{N_c}\alpha_i\mathsf{I}(\mathbf{s};\mathbf{y}|\mathbf{H},\mathbf{E}_i)+\mathsf{I}(\mathbf{E}_i;\mathbf{y}|\mathbf{H},\mathbf{s}).
\end{split}
\end{equation}
Obviously in (\ref{eq18}), the input distribution to the RF chains conditioned when $\mathbf{E}_i$ is activated, i.e., $f_{\mathbf{s}|\mathbf{E}_i}(\mathbf{s}|\mathbf{E}_i)$ not only affects the terms $\sum_{i=1}^{N_c}\alpha_i\mathsf{I}(\mathbf{s};\mathbf{y}|\mathbf{H},\mathbf{E}_i)$ but also the term $\mathsf{I}(\mathbf{E}_i;\mathbf{y}|\mathbf{H},\mathbf{s})$. It is certain that all $\{f_{\mathbf{s}|\mathbf{E}_i}(\mathbf{s}|\mathbf{E}_i)\}$ should be zero mean complex Gaussian distributed to maximize the terms $\sum_{i=1}^{N_c}\alpha_i\mathsf{I}(\mathbf{s};\mathbf{y}|\mathbf{H},\mathbf{E}_i)$, because $\mathsf{I}(\mathbf{s};\mathbf{y}|\mathbf{H},\mathbf{E}_i)$ is actually the mutual information between the input and output of the constant $N_{RF}\times N_r$ MIMO channel with channel matrix $\mathbf{HE}_i$ \cite{Yang2008}.
However, it is hard to strictly prove that the complex Gaussian distributed $f_{\mathbf{s}|\mathbf{E}_i}(\mathbf{s}|\mathbf{E}_i)$ can meanwhile maximize the term $\mathsf{I}(\mathbf{E}_i;\mathbf{y}|\mathbf{H},\mathbf{s})$. Therefore,  we next resort to discussing the capacity-achieving input distribution in the high SNR regime.

Based on the relationship
\begin{equation}
\mathsf{I}(\mathbf{E}_i;\mathbf{y}|\mathbf{H},\mathbf{s})\leq \mathsf{H}(\mathbf{E}_i)=\mathsf{H}(\boldsymbol{\alpha})=-\sum_{i=1}^{N_c}\alpha_i\log_2 \alpha_i,
\end{equation}  we obtain an upper bound of the capacity as
\begin{equation}\label{UB}
\begin{split}
\overline{\mathsf{C}_{\mathbf{H}}}
&=\max_{\boldsymbol{\alpha},f_{\mathbf{s}|\mathbf{E}_i}(\mathbf{s}|\mathbf{E}_i)}\sum_{i=1}^{N_c}\alpha_i\mathsf{I}(\mathbf{s};\mathbf{y}|\mathbf{H},\mathbf{E}_i)-\sum_{i=1}^{N_c}\alpha_i\log_2 \alpha_i.
\end{split}
\end{equation}
Observing (\ref{UB}), we find that $f_{\mathbf{s}|\mathbf{E}_i}(\mathbf{s}|\mathbf{E}_i)$ only affects the terms $\sum_{i=1}^{N_c}\alpha_i\mathsf{I}(\mathbf{s};\mathbf{y}|\mathbf{H},\mathbf{E}_i)$. Therefore, the zero mean complex Gaussian distributed  $f_{\mathbf{s}|\mathbf{E}_i}(\mathbf{s}|\mathbf{E}_i)$ can achieve  $\overline{\mathsf{C}_{\mathbf{H}}}$, and we obtain
\begin{equation}\label{UB1}
\begin{split}
\overline{\mathsf{C}_{\mathbf{H}}}
=\max_{\boldsymbol{\alpha},\{\mathbf{Q}_i\}}\sum_{i=1}^{N_c}\alpha_i\log_2\det\left(\mathbf{D}_i\right)-\sum_{i=1}^{N_c}\alpha_i\log_2\alpha_i,
\end{split}
\end{equation}
where 
\begin{equation}
\mathbf{D}_i=\mathbf{I}_{N_r}+\gamma\mathbf{H}\mathbf{E}_i\mathbf{Q}_i\mathbf{E}_i^H\mathbf{H}^H,
\end{equation}
and $\mathbf{Q}_i$ denotes the variance matrix of $\mathbf{s}|\mathbf{E}_i$ and the power constraint can be rewritten as $\sum_{i=1}^{N_c}\alpha_i\tr(\mathbf{Q}_i)\leq 1$.
Since $\mathbf{s}|\mathbf{E}_i\sim\mathcal{CN}(\mathbf{0},\mathbf{Q}_i)$, the PDF of the input vector $\mathbf{x}$ to the MIMO channel $\mathbf{H}$ can be expressed as
\begin{equation}\label{PDFx}
\begin{split}
f_{\mathbf{x}}(\mathbf{x})&=\sum_{i=1}^{N_c}\alpha_i f_{\mathbf{E}_i\mathbf{s}}(\mathbf{E}_i\mathbf{s})
\\&=\sum_{i=1}^{N_c}\alpha_i\frac{1}{\pi^{N_t}\det(\mathbf{E}_i\mathbf{Q}_i\mathbf{E}_i^H)}\exp\left[-\mathbf{x}^H(\mathbf{E}_i\mathbf{Q}_i\mathbf{E}_i^H)^{-1}\mathbf{x}\right].
\end{split}
\end{equation}
The $f_{\mathbf{x}}(\mathbf{x})$ in (\ref{PDFx}) is indeed a Gaussian mixture distribution with mixture coefficients $\{\alpha_i\}$ and we have the following theorem:

\begin{theorem}
Given $\boldsymbol{\alpha}^*,\{\mathbf{Q}_i^*\}$ as the solutions to attain $\overline{\mathsf{C}_{\mathbf{H}}}$,
the Gaussian mixture input in (\ref{PDFx}) with $\mathbf{a}=\mathbf{a}^*$ and $\{\mathbf{Q}_i\}=\{\mathbf{Q}^*_i\}$ is capacity-achieving in the high SNR regime. The capacity expression in the high SNR regime can be approximate to be
\begin{equation}\label{CHa}
\mathsf{C}_\mathbf{H}\approx\sum_{i=1}^{N_c}\alpha_i^*\log_2\det\left(\mathbf{D}_i^*\right)-\sum_{i=1}^{N_c}\alpha_i^*\log_2\alpha_i^*,
\end{equation}
where 
\begin{equation}
\mathbf{D}_i^*=\mathbf{I}_{N_r}+\gamma\mathbf{H}\mathbf{E}_i\mathbf{Q}_i^*\mathbf{E}_i^H\mathbf{H}^H.
\end{equation}
\end{theorem}
\begin{IEEEproof}
Using the Gaussian mixture with PDF $f_{\mathbf{x}}(\mathbf{x})$ in (\ref{PDFx}) as inputs, one can derive the mutual information between $\mathbf{x}$ and $\mathbf{y}$ for any $\boldsymbol{\alpha}$ and $\{\mathbf{Q}_i\}$ based on the entropy analysis of Gaussian mixture model as \cite{Huber2008,Ibrahim2016}
\begin{equation}\label{MI}
\begin{split}
&\mathsf{R}\left(\boldsymbol{\alpha},\{\mathbf{Q}_i\}\right)\\&=\mathsf{I}(\mathbf{x};\mathbf{y}|\mathbf{H})
\\&=\mathsf{H}(\mathbf{y})-\mathsf{H}(\mathbf{y}|\mathbf{H},\mathbf{x})
\\&=\mathbb{E}[-\log_2 f(\mathbf{y})]-N_r\log_2(\pi e)
\\&=-\int_{\mathbb{C}^{N_r}}\log_2 f(\mathbf{y})\sum_{i=1}^{N_c}\alpha_i f_i(\mathbf{y}) dy-N_r\log_2(\pi e),
\end{split}
\end{equation}
where 
\begin{equation}
f(\mathbf{y})=\sum_{i=1}^{N_c}\alpha_i f_i(\mathbf{y}),
\end{equation}
and
\begin{equation}
f_i(\mathbf{y})=\frac{1}{\pi^{N_r}\det(\mathbf{D}_i)}\exp\left(-\mathbf{y}^H\mathbf{D}_i^{-1}\mathbf{y}\right).
\end{equation}

Based on the expression of $\mathsf{R}\left(\boldsymbol{\alpha},\{\mathbf{Q}_i\}\right)$ in (\ref{MI}) and the constraint $N_r>N_{RF}$, we can obtain
\begin{equation}\label{eq24}
\begin{split}
\lim_{\gamma\rightarrow +\infty}\mathsf{R}\left(\boldsymbol{\alpha}^*,\{\mathbf{Q}_i^*\}\right)
&=\sum_{i=1}^{N_c}\alpha_i^*\log\det\left(\mathbf{D}_i^*\right)-\sum_{i=1}^{N_c}\alpha_i^*\log_2\alpha_i^*\\&=\overline{\mathsf{C}_{\mathbf{H}}},
\end{split}
\end{equation}
the proof of which is available in Appendix A.

According to information theory and the definition of $\overline{\mathsf{C}_{\mathbf{H}}}$,  we have 
\begin{equation}\label{eq25}
\mathsf{R}\left(\boldsymbol{\alpha}^*,\{\mathbf{Q}_i^*\}\right)\leq \mathsf{C}_{\mathbf{H}}\leq \overline{\mathsf{C}_{\mathbf{H}}}.
\end{equation}
Summarizing (\ref{eq24}) and (\ref{eq25}), it is proven that
\begin{equation}
\lim_{\gamma\rightarrow +\infty}\mathsf{R}\left(\boldsymbol{\alpha}^*,\{\mathbf{Q}_i^*\}\right)=\mathsf{C}_{\mathbf{H}}.
\end{equation}
\end{IEEEproof}
Then, adopting the Lagrange multiplier method, we derive the optimal $\boldsymbol{\alpha}^*$ and $\{\mathbf{Q}_i^*\}$ as given in the following theorem.
\begin{theorem}
If $\textrm{rank}(\mathbf{HE}_1)=\cdots=\textrm{rank}(\mathbf{HE}_{N_c})=m_s$,
 the optimal $\{\mathbf{Q}_i^*\}$ can be given by
\begin{equation}
\mathbf{Q}_i^*=\mathbf{V}_i{\mathbf{\Sigma}_i}^*{\mathbf{\Sigma}_i^*}^H\mathbf{V}_i^H,
\end{equation}
where $\mathbf{V}_i\in\mathbb{C}^{N_{RF}\times m_s}$ is the right-singular matrix of $\mathbf{HE}_i$ for channel parallelization;  $\mathbf{\Sigma}_i^*=\diag(\sqrt{\sigma_{i1}^*},\cdots,\sqrt{\sigma_{im_s}^*})$ is a diagonal water-filling power allocation matrix to satisfy $\tr{(\mathbf{Q}_i^*)}=1$ and
\begin{equation}
\sigma_{i,j}^*=\left(\frac{1}{\xi_i\ln2}-\frac{1}{\gamma\lambda_{ij}^2}\right)^+,~j=1,\cdots,m_s
\end{equation}
in which  $\lambda_{ij}^2$ is the $j$-channel gain of $\mathbf{HE}_i$ after channel parallelization  and $\xi$  satisfies
\begin{equation}
\sum_{j=1}^{m_s}\left(\frac{ 1}{\xi_{i}\ln2}-\frac{1}{\gamma\lambda_{ij}^2}\right)^+=1.
\end{equation}
 The optimal $\boldsymbol{\alpha}^*$ can be given by
\begin{equation}
\alpha_i^*=\frac{\det(\mathbf{D}_i^*)}{\sum_{i=1}^{N_c}\det(\mathbf{D}_i^*)},~i=1,\cdots,N_c.
\end{equation}
\end{theorem}
\begin{IEEEproof}
See Appendix B.
\end{IEEEproof}
When there is only a single RF chain, $\det(\mathbf{D}_i^*)\approx\lambda_{i1}^2$ and $\alpha_i^*\approx\frac{\lambda_{i1}^2}{\sum_{i=1}^{N_c}\lambda_{i1}^2}$. That is, the Gaussian mixture coefficients representing channel activation probabilities are approximately proportional to the channel gains and the channel with a higher channel gain will be activated with a higher probability.
\begin{theorem}
The capacity in the high SNR regime can be approximately to be 
\begin{equation}\label{CH}
\mathsf{C}_{\mathbf{H}}\approx\log_2 \sum_{i=1}^{N_c}\det{(\mathbf{D}_i^*)},
\end{equation}
\end{theorem}
\begin{IEEEproof}
Substituting the optimal $\boldsymbol{\alpha}^*$  and $\{\mathbf{Q}_i^*\}$ in Theorem 2 into the capacity approximation (\ref{CHa})  in Theorem 1, one can easily obtain the expression of capacity in (\ref{CH}) in the high SNR regime.
\end{IEEEproof}
For the special case with a single RF chain, $\mathsf{C}_{\mathbf{H}}\approx\log_2 \sum_{i=1}^{N_c}{\lambda_{i1}^2\gamma}$, which achieves the MRC gain of multiple channels.

By replacing $\mathbf{H}$, $\mathbf{E}_i$, $\mathbf{x}$, $\mathbf{y}$ with $\mathbf{\Lambda}$, $\mathbf{E}_i'$, $\mathbf{x'}$ , $\mathbf{y'}$ in all derivations, one can easily extend the MIMO capacity analysis to systems employing beamformers as connectors. For brevity, we omit it.

\section{Examples to Gain Insights}
\begin{figure}[t]
  \centering
  \includegraphics[width=0.42\textwidth]{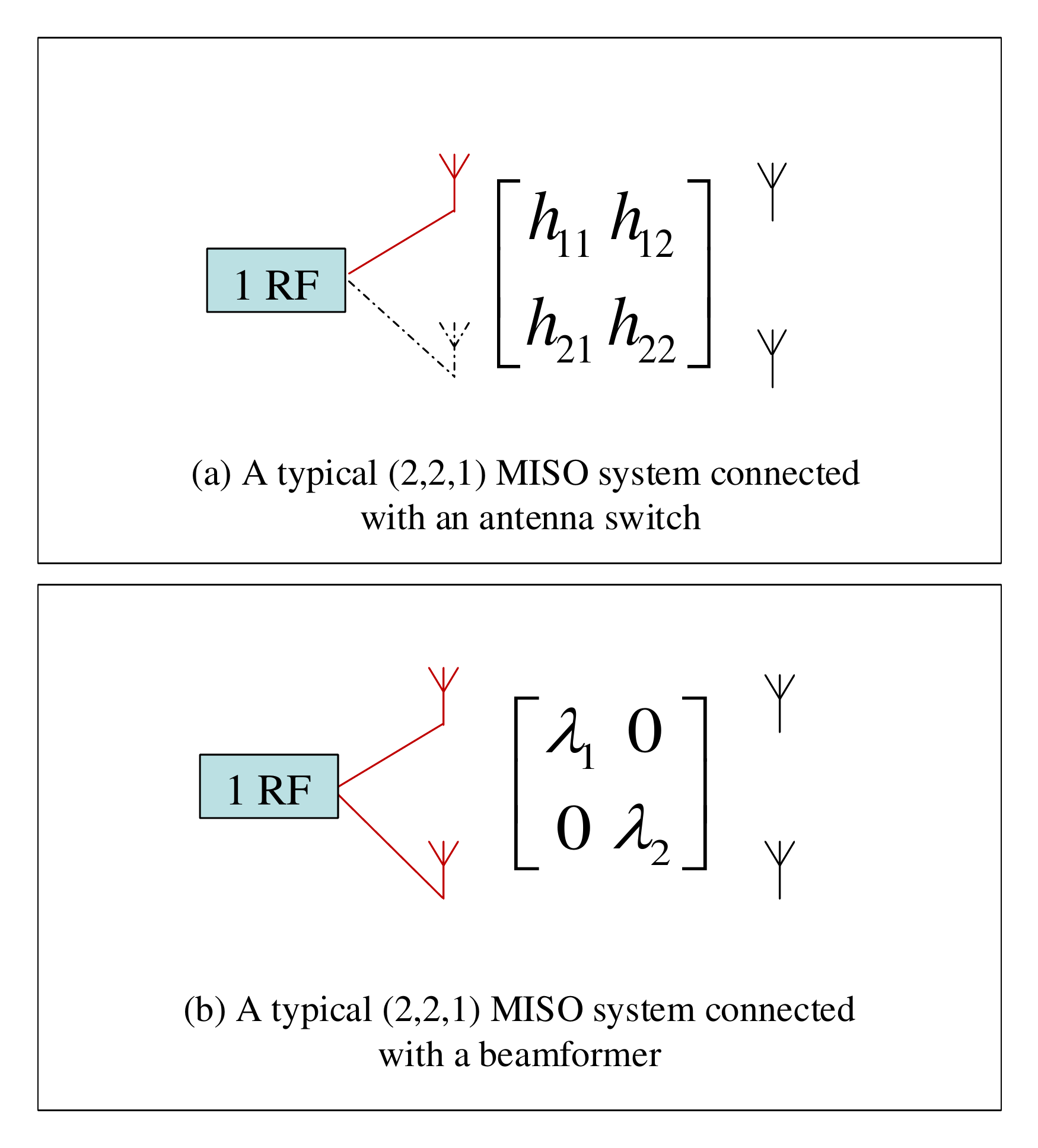}\\
 \caption{Typical $(2,2,1)$ MIMO systems.}
  \label{TIT_example}
\end{figure}

The Gaussian mixture distribution input can be realized by the newly emerging techniques: NUSM in \cite{Liu2016} and NUBM in \cite{Guo2019}, where the indices of antenna/beamspace combinations are used to convey information. In the high SNR regime, NUSM/NUBM activating the $i$-th antenna/beamspace combination with probability $\boldsymbol{\alpha}^*$ and sending zero mean variance $\mathbf{Q}_i^*$ distributed $\mathbf{s}|\mathbf{E}_i$ (or $\mathbf{s}|\mathbf{E}_i'$) achieves the capacity.  To show the superiority of NUSM/NUBM, we investigate the comparison with BAS/BBS and USM/UBM in typical  $(2,2,1)$ MIMO systems, as depicted in Figs. \ref{TIT_example}(a)-(b). In the $(2,2,1)$ MIMO system with an antenna switch as the connector, BAS, USM and NUSM are compared and the channel matrix is given to be
\begin{equation}
\mathbf{H}_2=
\begin{bmatrix}
h_{11}&h_{12}\\
h_{21}&h_{22}
\end{bmatrix}.
\end{equation}
 In the $(2,2,1)$ MIMO system with a beamformer as the connector, BBS, UBM and NUBM are compared and the channel matrix is given by
\begin{equation}
\mathbf{H}_2=\mathbf{U}\begin{bmatrix}
\lambda_1&0\\
0&\lambda_2
\end{bmatrix}\mathbf{V}^H.
\end{equation}
For both comparisons, there are two transmit antennas/beamspaces available whereas there is only a single RF chain. For unification, we introduce $\{g_i\}$ to represent the channel gains for both cases, i.e., $g_i=|h_{1i}|^2+|h_{2i}|^2$ or $g_i=\lambda_i^2$ for $i=1,2$.
\subsection{Applicable Techniques}
\subsubsection{BAS/BBS}
For BAS/BBS, the antenna/beamspace with the superior capacity is selected to convey information and the achievable capacity can be written as
\begin{equation}\label{BASBBS}
\begin{split}
\mathsf{C}_{\textrm{BAS/BBS}}&=\log_2(1+\gamma\max\{g_1,g_2\})\\&\approx \log_2(\gamma\max\{g_1,g_2\}),
\end{split}
\end{equation}
where the approximation is accurate in the high SNR regime.
\subsubsection{USM/UBM}
USM/UBM utilizes equal-probability antenna/beamspace activation to carry information, i.e., $\alpha_1=\alpha_2=\frac{1}{2}$. Without power allocation, the achievable capacity in the high SNR regime can be approximate as
\begin{equation}\label{USMUBM}
\begin{split}
\mathsf{C}_{\textrm{USM/UBM}}^{\textrm{No~PA}}&\approx\frac{1}{2}\log_2(1+\gamma g_1)+\frac{1}{2}\log_2(1+\gamma g_2)+1\\
&\approx\frac{1}{2}\log_2(\gamma g_1)+\frac{1}{2}\log_2(\gamma g_2)+1\\
&=\log_2(2\gamma\sqrt{g_1g_2}).
\end{split}
\end{equation}
It is should mentioned that CSIT is not a necessity  for USM/UBM. That is, USM/UBM can work without CSIT. In the case with CSIT, the transmitter can adopt water-filling power allocation for capacity maximization \cite{Shi2017} and the corresponding capacity in the high SNR regime can be given by
\begin{equation}
\begin{split}
\mathsf{C}_{\textrm{USM/UBM}}^{\textrm{PA}}&=\frac{1}{2}\log_2(1+\gamma \beta_1g_1)+\frac{1}{2}\log_2(1+\gamma\beta_2 g_2)+1\\
&\approx\frac{1}{2}\log_2(\gamma\beta_1 g_1)+\frac{1}{2}\log_2(\gamma \beta_2g_2)+1\\
&=\log_2(2\gamma\sqrt{\beta_1g_1\beta_2g_2}),
\end{split}
\end{equation}
where $\beta_i$ can be given as
\begin{equation}
\beta_{i}=\left(\frac{1}{\eta\ln2}-\frac{1}{\gamma g_i}\right)^+,~i=1,2
\end{equation}
and
\begin{equation}
\sum_{i=1}^{2}\left(\frac{ 1}{\eta \ln2}-\frac{1}{\gamma g_{i}}\right)^+=2.
\end{equation}
As $\gamma$ goes to infinity, $\beta_1\approx\beta_2\approx 1$ and
\begin{equation}
\mathsf{C}_{\textrm{USM/UBM}}^{\textrm{No~PA}}\approx\mathsf{C}_{\textrm{USM/UBM}}^{\textrm{PA}}.
\end{equation}
\subsubsection{NUBM/NUBM}
According to Theorem 2, NUSM/NUBM utilizes the optimal $\alpha_1^*\approx\frac{g_1}{g_1+g_2}$, $\alpha_2^*\approx\frac{g_2}{g_1+g_2}$ for achieving the capacity in the high SNR regime. The capacity in the high SNR regime can be approximately written as
\begin{equation}\label{NUSMNUBM}
\mathsf{C}_{\textrm{NUSM/NUBM}}\approx\log_2(g_1\gamma+g_2\gamma).
\end{equation}
\subsection{Remark and Insight}
Observing (\ref{NUSMNUBM}), we find that NUSM/NUBM can achieve MRC gain of two channels.
Comparing (\ref{BASBBS}) with (\ref{USMUBM}) and (\ref{NUSMNUBM}), we find that USM/UBM in 
\cite{Mesleh2008,Yang2008,Renzo2014,Guo2016,Guo2017,Ibrahim2016,Basnayaka2016,Faregh2016,Narasimhan2016,Younis2017,Younis2018} outperform BAS/BBS only if $2\sqrt{g_1g_2}>\max\{g_1,g_2\}$ in the high SNR regime, while NUSM/NUBM in 
\cite{Liu2016,Guo2019} is certain to outperform BAS/BBS and USM/UBM, because $g_1+g_2\geq\max\{g_1,g_2\}$ and $g_1+g_2\geq2\sqrt{g_1g_2}$. These findings can provide significant insights and can offer useful design guidelines for future MIMO systems requiring expensive and power-hungry RF chains, especially for mmWave MIMO and Terahertz MIMO. For instance, in antenna selection-based systems, feeding back the optimal antenna activation probabilities  to  enable NUSM  is better than only feeding the optimal antenna index back.

\section{Simulation and Discussion}
In this section, we demonstrate the spectrum efficiency comparison over constant channels and ergodic spectrum efficiency comparison over $100$ independent Rayleigh channel realizations.
\subsection{Spectral Efficiency Comparison Over Constant Channels}
First, we simulate the spectral efficiencies of  $(2,2,1)$ MIMO systems employing various techniques over the following two channel realizations:
\begin{figure}[t]
  \centering
  \includegraphics[width=0.5\textwidth]{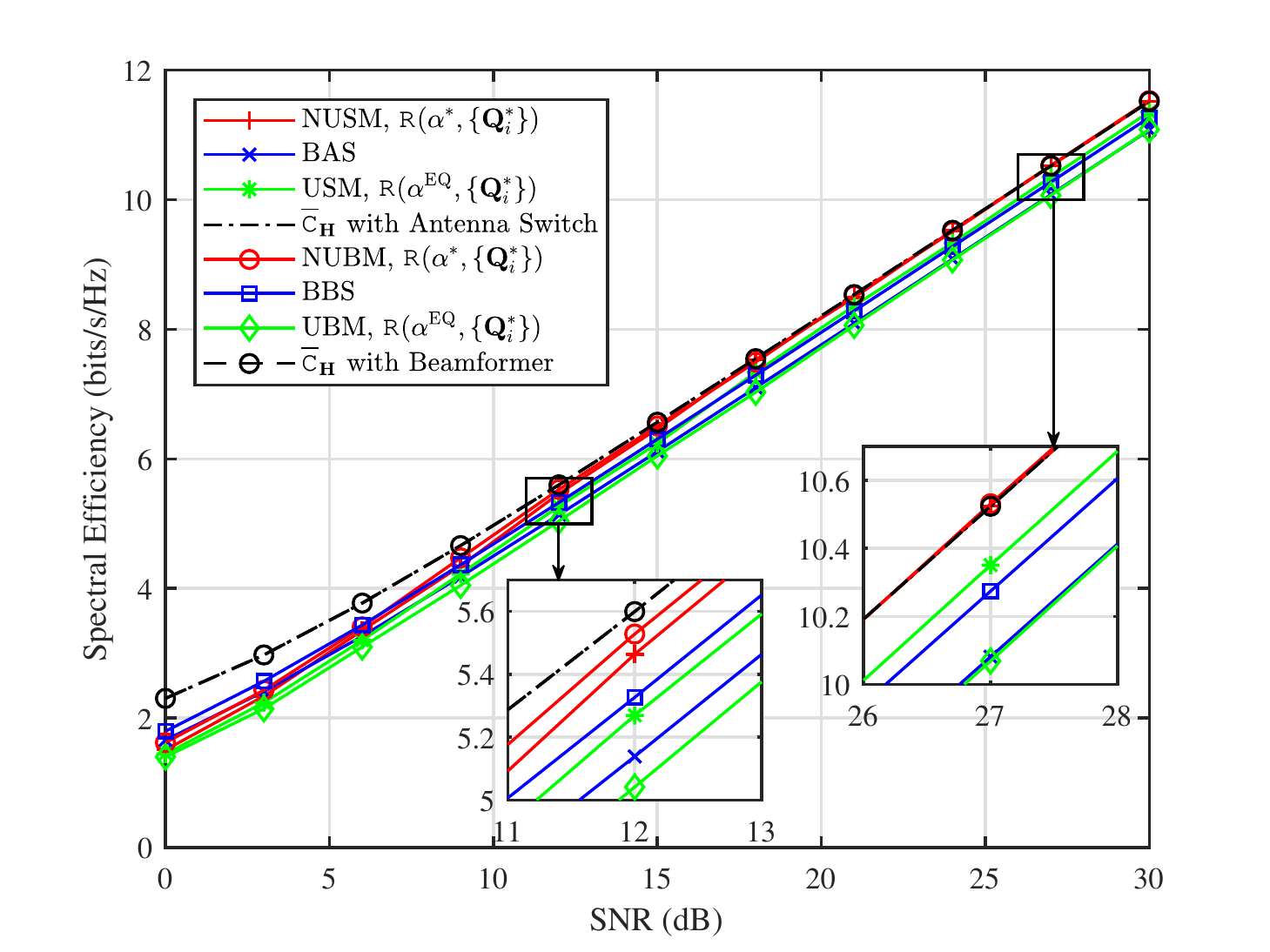}\\
 \caption{Spectral efficiency comparison in $(2,2,1)$ MIMO systems over the channel $\mathbf{H}_{2\times 2}^{(1)}$.}
  \label{result0a}
\end{figure}
\begin{figure}[t]
  \centering
  \includegraphics[width=0.5\textwidth]{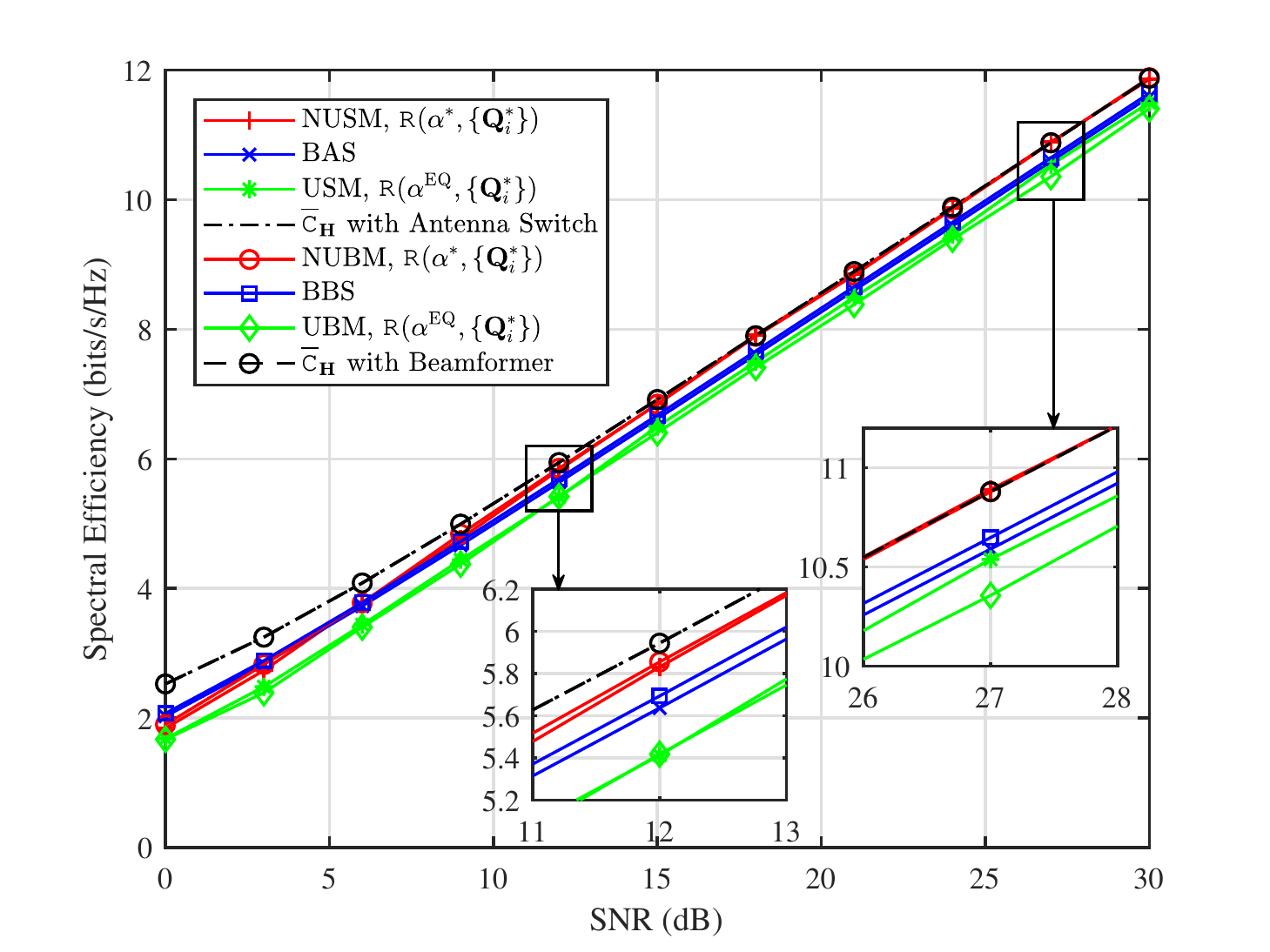}\\
 \caption{Spectral efficiency comparison in  $(2,2,1)$ MIMO systems over the channel $\mathbf{H}_{2\times 2}^{(2)}$.}
  \label{result0b}
\end{figure}\begin{equation}
\mathbf{H}_{2\times 2}^{(1)}=  
\begin{bmatrix}
-0.0062 + 0.8531\imath  &-0.3000 - 0.4998\imath\\
   0.1650 + 0.1395\imath & -1.0391 + 0.8601\imath
\end{bmatrix},
\end{equation}
and
\begin{equation}
\mathbf{H}_{2\times 2}^{(2)}=  
\begin{bmatrix}
 0.1901 + 0.1127\imath  &  -1.5972 - 0.3066\imath  \\
   0.6484 - 0.4623\imath  &   0.6096 + 0.2423\imath  
\end{bmatrix},
\end{equation}
where $\imath=\sqrt{-1}$. 
\begin{figure*}
\begin{equation}\label{H53}
\mathbf{H}_{3\times 5}=
\begin{bmatrix}
   0.2248 + 0.9300\imath & -1.6193 + 0.1124\imath &  0.7983 + 1.4626\imath  & 0.7101 - 0.5903\imath & -1.1416 - 0.5433\imath\\
   0.6925 + 0.2182\imath & -0.4773 - 0.7437\imath&   0.5020 + 0.0908\imath  & 0.5070 + 0.3351\imath&  -0.7329 + 0.3995\imath\\
  -1.2944 - 1.2016\imath &  0.2182 - 0.2740\imath &  1.5437 + 0.1045\imath  & 0.5398 - 0.6127\imath &  0.1665 - 0.0869\imath
\end{bmatrix}.
\end{equation}
\end{figure*}
\begin{figure}[t]
  \centering
  \includegraphics[width=0.5\textwidth]{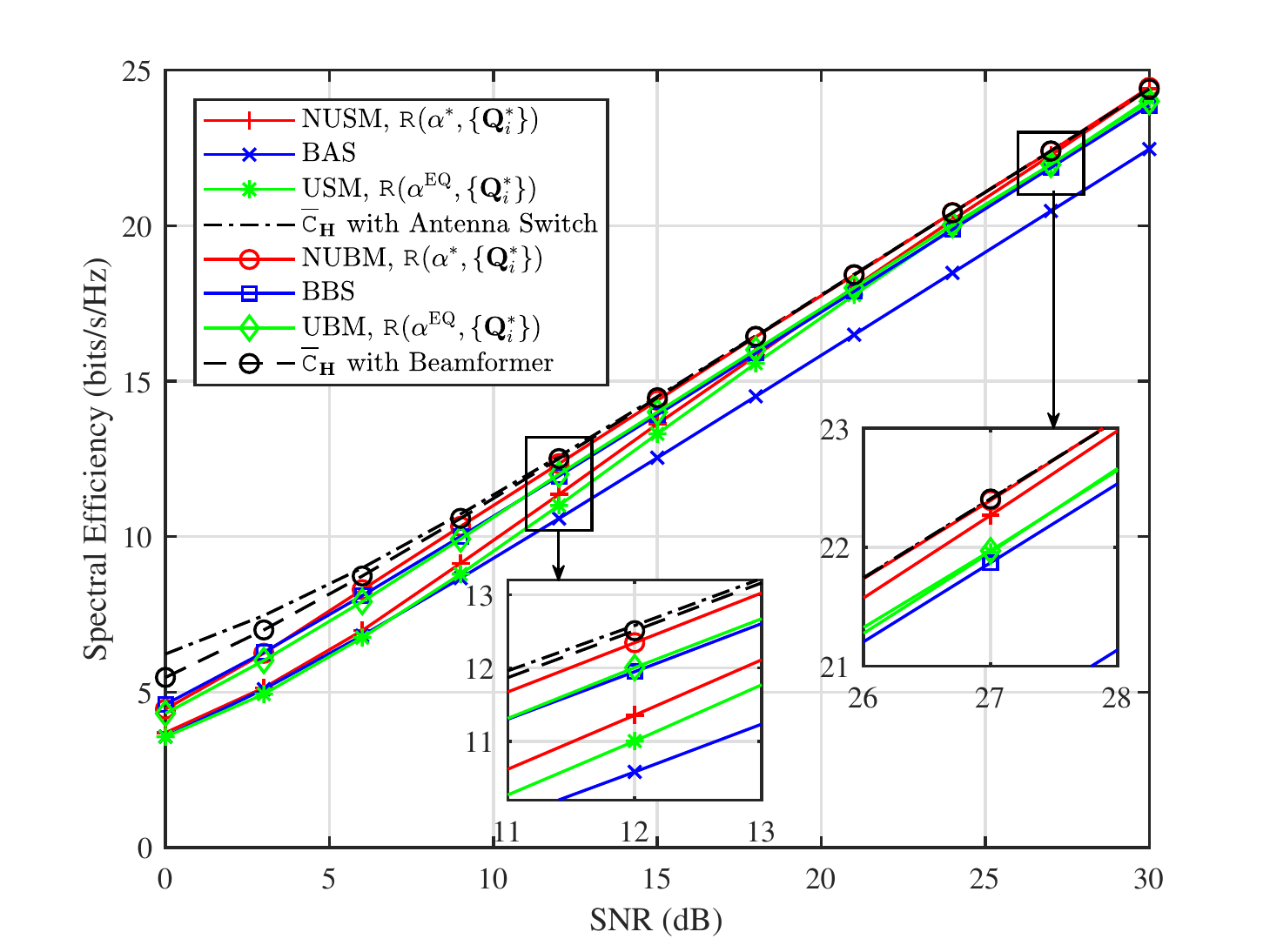}\\
 \caption{Spectral efficiency comparison in $(5,3,2)$ MIMO systems over the channel $\mathbf{H}_{5\times 3}$.}
  \label{result2}
\end{figure}
\begin{figure}[t]
  \centering
  \includegraphics[width=0.5\textwidth]{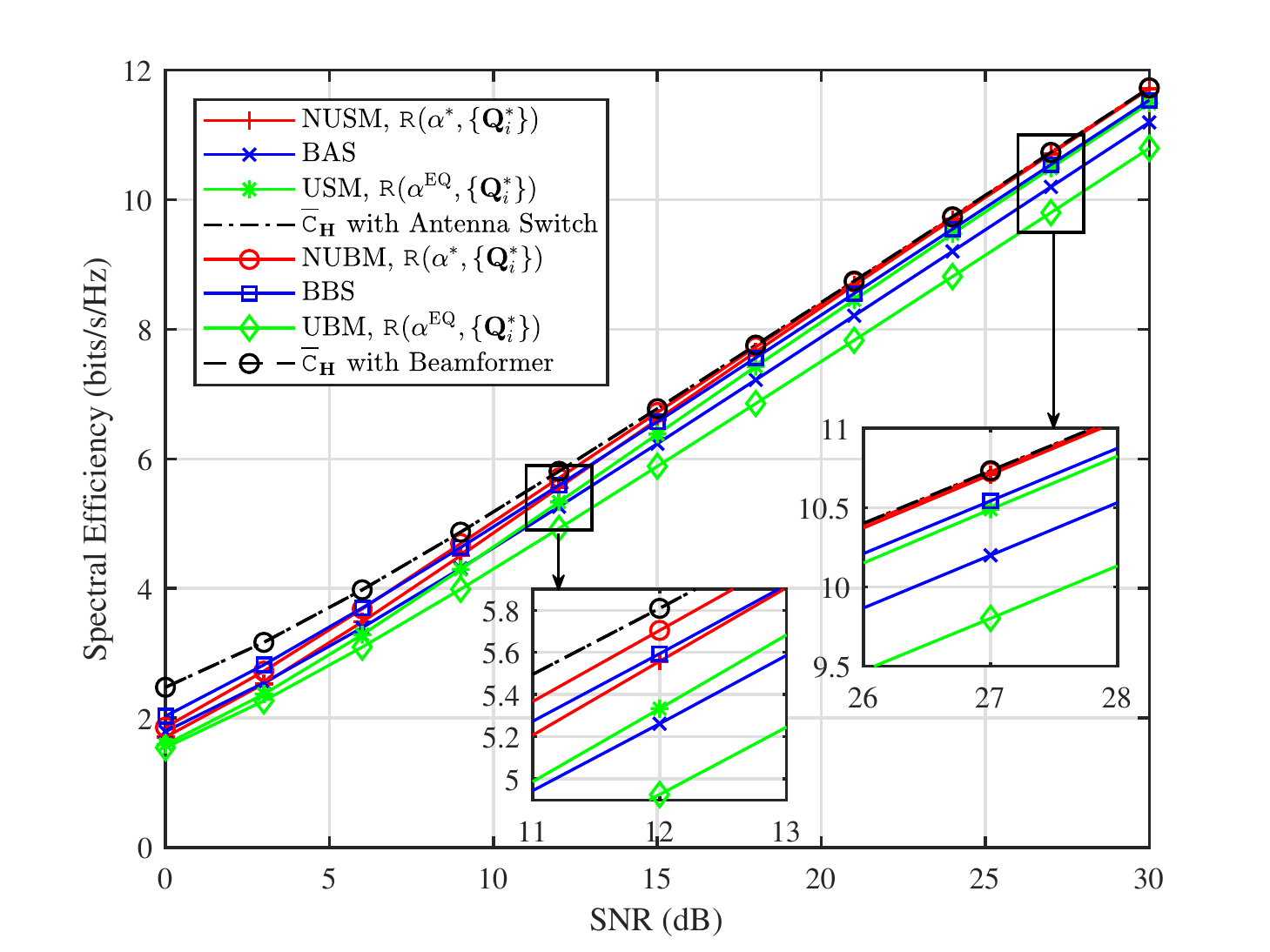}\\
 \caption{Ergodic spectral efficiency comparison in $(2,2,1)$ MIMO systems over independent Rayleigh channels.}
  \label{result0c}
\end{figure}
\begin{figure}[t]
  \centering
  \includegraphics[width=0.5\textwidth]{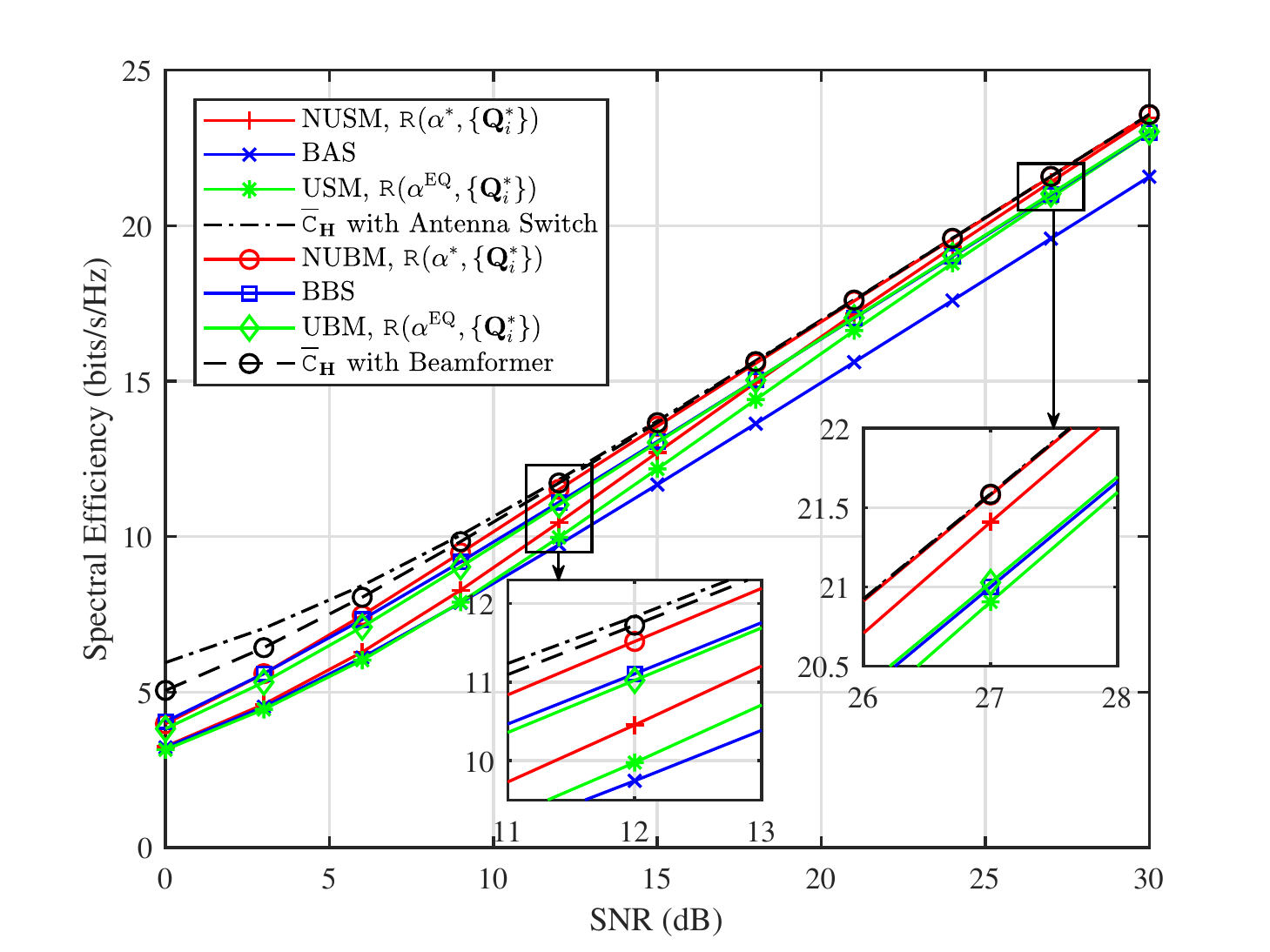}\\
 \caption{Ergodic spectral efficiency comparison in $(5,3,2)$ MIMO systems over independent Rayleigh channels.}
  \label{result4}
\end{figure}

The simulation results are illustrated in Figs. \ref{result0a} and \ref{result0b}, respectively.  Observing these two figures, we find that the capacity upper bounds in the high SNR regime for systems adopting antenna switches and beamformers as connectors are close in the all depicted SNR regime.
Both NUSM and NUBM are capacity-achieving in the high SNR regime. Besides the high SNR regime, we also demonstrate the comparisons in the medium SNR regime. Numerical results show that NUBM slightly outperforms NUSM, because NUSM activates one antenna for each time slot while NUBM activates two and as a result, NUSM achieves a smaller beamforming gain. Compared with USM/UBM employing equal-probability antenna/beamspace activation (i.e.,$\boldsymbol{\alpha}^{\mathrm{EQ}}$ in the figures) and BAS/BBS employing the best antenna/beamspace, NUSM/NUBM shows the superiority in the medium-to-high SNR regime. The improvement in spectrum efficiency is appealing. For instance, according to the numerical results over $\mathbf{H}_{2\times 2}^{(1)}$ as depicted in Fig. \ref{result0a}, system spectral efficiency increases from $5$ bits/s/Hz to $5.5$ bits/s/Hz by $10\%$ at a medium SNR of $12$ dB.  As discussed in Section IV, NUSM/NUBM always outperforms BAS/BBS in the high SNR regime and in contrast  USM/UBM is superior to BAS/BBS only under the channel condition $2\sqrt{g_1g_2}>\max\{g_1,g_2\}$. This is also validated by the numerical results. Specifically, for the given channel realization $\mathbf{H}_{2\times 2}^{(1)}$, USM is superior to BAS, while for the given channel realization $\mathbf{H}_{2\times 2}^{(2)}$, USM is inferior to BAS. Additionally, for both realizations, UBS is inferior to BBS. Observing the simulations results in the low SNR regime, we find that NUSM/NUBM with optimized $\boldsymbol{\alpha}^*$ and $\{\mathbf{Q}_i^*\}$ achieves lower spectral efficiency than BAS/BBS, which indicates that NUSM/NUBM with $\boldsymbol{\alpha}^*$ and $\{\mathbf{Q}_i^*\}$ cannot achieve the system capacity in the low SNR regime.

To demonstrate the spectral efficiency comparison in MIMO systems with multiple RF chains, we simulate the spectral efficiencies of $(5,3,2)$ MIMO systems employing various techniques over a channel realization $\mathbf{H}_{3\times 5}$ given in (\ref{H53}).  The simulation results are demonstrated in Fig. \ref{result2}. From the comparison in the high SNR regime at an SNR of $27$ dB, it is observed that NUSM/NUBM outperforms other schemes and approximately achieves the capacity at the SNR of $27$ dB. From the comparison in the middle SNR regime at an SNR of $12$ dB, we see that systems with beamformers as connectors outperforms that with antenna switches as connectors. That is, BBS/UBM/NUBM outperforms BAS/USM/NUSM. The reason is that BBS/UBM/NUBM benefits from a higher beamforming gain by activating all antennas at each time slot and the beamforming gain is helpful in improving the system spectrum efficiency in the low-to-medium SNR regime.

\subsection{Ergodic Spectral Efficiency Comparison Over Independent Rayleigh Channels}
Besides the comparisons over some specific channel realizations, we also include the ergodic spectral efficiency comparisons, where the simulation results are averaged over $100$ channel realizations. The comparison results in $(2,2,1)$ and $(5,3,2)$ MIMO systems are illustrated in Figs. \ref{result0c} and \ref{result4}, respectively. Numerical results in Fig. \ref{result0c} shows that NUSM/NUBM outperforms other schemes in the high SNR regime in $(2,2,1)$ MIMO systems over independent channels. In the medium SNR regime at an SNR of $12$ dB, we have the priority order as $\textrm{NUBM}\succ \textrm{BBS}\succ \textrm{NUSM}\succ \textrm{USN}\succ \textrm{BAS}\succ \textrm{UBM}$. In $(5,3,2)$ MIMO systems over independent Rayleigh channels, we observe the similar numerical results in the high SNR regime that NUBM is  the best; NUBM slightly outperforms NUSM at an SNR of $27$ dB; both of them considerably outperforms others. In the medium SNR regime at an SNR of $12$ dB, the priority order is: $\textrm{NUBM}\succ \textrm{BBS}\succ \textrm{UBM}\succ \textrm{NUSM}\succ \textrm{USM}\succ \textrm{BAS}$.

\section{Conclusion}
In this work, we characterized the MIMO capacity with insufficient RF chains as the maximum mutual information given any vector input subject to not only an average power constraint but also a sparsity constraint. We proved that the capacity-achieving input distribution is a Gaussian mixture distribution in the high SNR regime. We derived that the optimal mixture coefficients and covariances matrices of the Gaussian mixtures. We obtained the expression of the capacity. For the special case with a single RF chain,  the optimal mixture coefficients are approximately proportional to the channel gains and the MIMO capacity is indeed the MRC of multiple channels. Compared previous solutions to select the best antenna/beamspace for conveying information, which achieves the selection gain of multiple channels, the diversity  can be greatly improved. Also, we investigated the capacity-achieving techniques for MIMO systems with reduced RF chains: NUSM and NUBM and compared them with BAS/BBS as well as the widely investigated uniform spatial/beamspace modulation (USM/UBM). Both analytical and numerical results showed that NUSM/NUBM considerably outperforms BAS/BBS and USM/UBM in the medium-to-high SNR regime.

\section{Acknowledgment}
We thank Prof. Ahmed Sultan-Salem and Prof. Peng Zhang for the helpful discussions.

\appendices
\section{Proof of Equation (\ref{eq24})}
\begin{IEEEproof}
According to \cite{He2018}, the asymptotic spectral efficiency can be written as
\begin{equation}\label{eq24a}
\begin{split}
&\lim_{\gamma\rightarrow +\infty}\mathsf{R}\left(\boldsymbol{\alpha}^*,\{\mathbf{Q}_i^*\}\right)\\&=-\sum_{i=1}^{N_c}\alpha_i^*\log_2\left(\sum_{j=1}^{N_c }\frac{\alpha_j^*}{\det(\mathbf{D}^*_i+\mathbf{D}_j^*)}\right)+N_r,
\\&=-\sum_{i=1}^{N_c}\alpha_i^*\log\left(\frac{\alpha_i^*}{\det(2\mathbf{D}^*_i)}+\sum_{j\neq i}^{N_c }\frac{\alpha_j^*}{\det(\mathbf{D}^*_i+\mathbf{D}_j^*)}\right)+N_r.
\end{split}
\end{equation}
Since the term $\det(\mathbf{D}^*_i+\mathbf{D}_j^*)\propto\gamma^{N_r-N_{RF}}$ (c.f., [14, Eq. 41]) and $N_r>N_{RF}$, thus
\begin{equation}\label{eq50}
\lim_{\gamma\rightarrow +\infty}\sum_{j\neq i}^{N_c }\frac{\alpha_j^*}{\det(\mathbf{D}^*_i+\mathbf{D}_j^*)}=0.
\end{equation}
Substituting (\ref{eq50}) into (\ref{eq24a}), we prove
\begin{equation}\label{eq24b}
\begin{split}
\lim_{\gamma\rightarrow +\infty}\mathsf{R}\left(\boldsymbol{\alpha}^*,\{\mathbf{Q}_i^*\}\right)
&=\sum_{i=1}^{N_c}\alpha_i^*\log\det\left(\mathbf{D}_i^*\right)-\sum_{i=1}^{N_c}\alpha_i^*\log_2\alpha_i^*\\&=\overline{\mathsf{C}_{\mathbf{H}}}.
\end{split}
\end{equation}
\end{IEEEproof}
\section{Proof of Theorem 2}
\begin{IEEEproof}
The solutions $\boldsymbol{\alpha}^*,\{\mathbf{Q}_i^*\}$ to attain $\overline{\mathsf{C}_{\mathbf{H}}}$ can be obtained by solving
\begin{equation}
\begin{split}
\max_{\boldsymbol{\alpha},\{\mathbf{Q}_i\}}&\sum_{i=1}^{N_c}\alpha_i\log_2\det(\mathbf{D}_i)-\sum_{i=1}^{N_c}\alpha_i\log_2\alpha_i\\
=\max_{\boldsymbol{\alpha},\{\mathbf{Q}_i\}}&\sum_{i=1}^{N_c}\alpha_i\log_2\det\left(\mathbf{I}_{N_r}+\gamma\mathbf{H}\mathbf{E}_i\mathbf{Q}_i\mathbf{E}_i^H\mathbf{H}^H\right)\\&-\sum_{i=1}^{N_c}\alpha_i\log_2\alpha_i\\
\mathrm{s.~t.:}& \sum_{i=1}^{N_c}\alpha_i\tr(\mathbf{Q}_i)=1,~\sum_{i=1}^{N_c}\alpha_i=1.
\end{split}
\end{equation}
For any feasible $\{\mathbf{Q}_i\}$, the Lagrange function of the problem in (\ref{OP}) to find the optimal $\boldsymbol{\alpha}$ can be formulated as
\begin{equation}\label{eqL1}
\begin{split}
J(\boldsymbol{\alpha},\mu)=\sum_{i=1}^{N_c}\alpha_i\left( \log_2\det(\mathbf{D}_i)-\log_2\alpha_i\right)-\mu\left(\sum_{i=1}^{N_c}\alpha_i-1\right).
\end{split}
\end{equation}
We take  partial derivations of the Lagrange function in (\ref{eqL1}) with respect to $\alpha_i$ and obtain a set of equations as
\begin{equation}
\log_2\det(\mathbf{D}_i)-\log_2 \alpha_i-1/\ln2 -\mu=0,~i=1,\cdots,N_c.
\end{equation}
Solving them, we have
\begin{equation}
\alpha_i=\frac{2^{-\mu}\det(\mathbf{D}_i)}{e},~i=1,\cdots,N_c.
\end{equation}
Since $\sum_{i=1}^{N_c}\alpha_i=1$, the optimal $\alpha_i^*$ for any feasible $\{\mathbf{Q}_i\}$ can be expressed as
 \begin{equation}
\alpha_i^{*}=\frac{\det(\mathbf{D}_i)}{\sum_{i=1}^{N_c}\det(\mathbf{D}_i)},~i=1,\cdots,N_c.
\end{equation}
Next, we need to find the optimal $\{\mathbf{Q}_i^*\}$.
According to the channel parallelization  in \cite{Telatar1999}, $\mathbf{Q}_i$ can be expressed as
\begin{equation}\label{OP}
\mathbf{Q}_i=\mathbf{V}_i{\mathbf{\Sigma}_i}{\mathbf{\Sigma}_i}^H\mathbf{V}_i^H,
\end{equation}
where $\mathbf{V}_i\in\mathbb{C}^{N_{RF}\times m_s}$ is the right-singular matrix of $\mathbf{HE}_i$ for channel parallelization; $\mathbf{\Sigma}_i=\diag(\sqrt{\sigma_{i1}},\cdots,\sqrt{\sigma_{im_s}})$ is a diagonal power allocation matrix. Let $\mathbf{\Lambda}_{i}=\diag(\lambda_{i1},\cdots,\lambda_{im_s})\in\mathbb{C}^{m_s\times m_s}$  be the diagonal matrix composed of the non-zeros singular values of  $\mathbf{HE}_i$. Based on these denotations, the optimization problem in (\ref{OP}) can be reformulated to be
\begin{equation}\label{eq47}
\begin{split}
\max_{\boldsymbol{\alpha},\{\mathbf{\Sigma}_i\}}&\sum_{i=1}^{N_c}\alpha_i\log_2\det\left(\mathbf{I}_{N_r}+\gamma\mathbf{\Lambda}_i\mathbf{\Sigma}_i\mathbf{\Sigma}_i^H\mathbf{\Lambda}_i^H\right)\\&-\sum_{i=1}^{N_c}\alpha_i\log_2\alpha_i\\
\mathrm{s.~t.:}& \sum_{i=1}^{N_c}\alpha_i\tr(\mathbf{\Sigma}_i\mathbf{\Sigma}_i^H)=1,~\sum_{i=1}^{N_c}\alpha_i=1.
\end{split}
\end{equation}
According to  \cite{Goldsmith2003}, we can approximately calculate the term $\log_2\det\left(\mathbf{I}_{N_r}+\gamma\mathbf{\Lambda}_i\mathbf{\Sigma}_i\mathbf{\Sigma}_i^H\mathbf{\Lambda}_i^H\right)$ by
 \begin{equation}\label{eq48}
\log_2\det\left(\mathbf{I}_{N_r}+\gamma\mathbf{\Lambda}_i\mathbf{\Sigma}_i\mathbf{\Sigma}_i^H\mathbf{\Lambda}_i^H\right)=m_s\log_2 \tr(\mathbf{\Sigma}_i\mathbf{\Sigma}_i^H)+q_i,
\end{equation} 
where $q_i$ is a constant related to the channel gains. For any feasible $\boldsymbol{\alpha}$, substituting (\ref{eq48}) into (\ref{eq47}), we obtain a new optimization problem  as
\begin{equation}\label{eq49}
\begin{split}
\max_{\mathbf{b}}&\sum_{i=1}^{N_c}\alpha_i\left[m_s\log_2 b_i+q_i-\log_2\alpha_i\right]\\
\mathrm{s.~t.:}& \sum_{i=1}^{N_c}\alpha_ib_i=1,~\sum_{i=1}^{N_c}\alpha_i=1.
\end{split}
\end{equation}
where $\mathbf{b}=[b_1,\cdots,b_{N_c}]^T$ and  $b_i=\tr(\mathbf{\Sigma}_i\mathbf{\Sigma}_i^H)$.
Similarly, we can formulate the Lagrange function of the problem to find the optimal $\mathbf{b}$ to be
\begin{equation}\label{eqL3}
\begin{split}
J(\mathbf{b},\zeta)
=\sum_{i=1}^{N_c}\alpha_i\left[ m_s\log b_i+q_i+\log \alpha_i\right]-\zeta\left(\sum_{i=1}^{|\mathcal{F}|}\alpha_ib_i-1\right).
\end{split}
\end{equation}
 By taking the  partial  derivation of the Lagrange function in (\ref{eqL3}) with respect to $b_i$,  we obtain a set of equations as
\begin{equation}
\frac{m_s\alpha_i}{b_i}-\zeta \alpha_i=0,~i=1,\cdots, N_c.
\end{equation}
By solving them, we have 
\begin{equation}
b_1=\cdots=b_{N_c}=\frac{m_s}{\zeta}.
\end{equation}
Since $\sum_{i=1}^{N_c}\alpha_ib_i-1=0$,  the optimal solution can be derived as
\begin{equation}
b_1^*=\cdots=b_{N_c}^*=1.
\end{equation}
Then the optimized problem in (\ref{eq47}) to optimize  $\mathbf{\Sigma}_i$ can be simplified to be
\begin{equation}
\begin{split}
\max_{\{\mathbf{\Sigma}_i\}}&\log_2\det\left(\mathbf{I}_{N_r}+\gamma\mathbf{\Lambda}_i\mathbf{\Sigma}_i\mathbf{\Sigma}_i^H\mathbf{\Lambda}_i^H\right)\\
\mathrm{s.~t.:}& \tr(\mathbf{\Sigma}_i\mathbf{\Sigma}_i^H)=1.
\end{split}
\end{equation}
This is a typical power allocation problem for paralleled MIMO channels, and the optimal solution is the water-filling solution as that given in Theorem 2. Summarizing all above, Theorem 2 is proved. 
\end{IEEEproof}



\begin{thebibliography}{10}
\providecommand{\url}[1]{#1}
\csname url@samestyle\endcsname
\providecommand{\newblock}{\relax}
\providecommand{\bibinfo}[2]{#2}
\providecommand{\BIBentrySTDinterwordspacing}{\spaceskip=0pt\relax}
\providecommand{\BIBentryALTinterwordstretchfactor}{4}
\providecommand{\BIBentryALTinterwordspacing}{\spaceskip=\fontdimen2\font plus
\BIBentryALTinterwordstretchfactor\fontdimen3\font minus
  \fontdimen4\font\relax}
\providecommand{\BIBforeignlanguage}[2]{{%
\expandafter\ifx\csname l@#1\endcsname\relax
\typeout{** WARNING: IEEEtran.bst: No hyphenation pattern has been}%
\typeout{** loaded for the language `#1'. Using the pattern for}%
\typeout{** the default language instead.}%
\else
\language=\csname l@#1\endcsname
\fi
#2}}
\providecommand{\BIBdecl}{\relax}
\BIBdecl

\bibitem{Guo2019a}
S.~Guo, H.~Zhang, and M.-S. Alouini, ``Rethink {MIMO} capacity with reduced
  {RF} chains,'' \emph{submitted to ITA workshop 2019}, 2019.

\bibitem{Shannon1948}
C.~E. Shannon, ``A mathematical theory of communication,'' \emph{Bell System
  Technical Journal}, vol.~27, no.~3, pp. 379--423, Oct. 1948.

\bibitem{Shannon1949}
------, ``Communication in the presence of noise,'' \emph{Proceedings of the
  IRE}, vol.~37, no.~1, pp. 10--21, 1949.

\bibitem{Telatar1999}
E.~Telatar, ``Capacity of multi-antenna {Gaussian} channels,'' \emph{European
  transactions on telecommunications}, vol.~10, no.~6, pp. 585--595, Nov. 1999.

\bibitem{Goldsmith2003}
A.~Goldsmith, S.~A. Jafar, N.~Jindal, and S.~Vishwanath, ``Capacity limits of
  {MIMO} channels,'' \emph{{IEEE} J. Sel. Areas Commun.}, vol.~21, no.~5, pp.
  684--702, Jun. 2003.

\bibitem{Marzett2010}
T.~L. Marzetta, ``Noncooperative cellular wireless with unlimited numbers of
  base station antennas,'' \emph{{IEEE} Trans. Wireless Commun.}, vol.~9,
  no.~11, pp. 3590--3600, Nov. 2010.

\bibitem{Larsson2014}
E.~G. Larsson, O.~Edfors, F.~Tufvesson, and T.~L. Marzetta, ``Massive {MIMO}
  for next generation wireless systems,'' \emph{{IEEE} Commun. Mag.}, vol.~52,
  no.~2, pp. 186--195, Feb. 2014.

\bibitem{Molisch2005}
A.~F. Molisch, M.~Z. Win, Y.-S. Choi, and J.~H. Winters, ``Capacity of {MIMO}
  systems with antenna selection,'' \emph{{IEEE} Trans. Wireless Commun.},
  vol.~4, no.~4, pp. 1759--1772, Jul. 2005.

\bibitem{Ayach2014}
O.~E. Ayach, S.~Rajagopal, S.~Abu-Surra, Z.~Pi, and R.~W. Heath, ``Spatially
  sparse precoding in millimeter wave {MIMO} systems,'' \emph{{IEEE} Trans.
  Wireless Commun.}, vol.~13, no.~3, pp. 1499--1513, Mar. 2014.

\bibitem{Dai2015}
L.~Dai, X.~Gao, J.~Quan, S.~Han, and C.~I, ``Near-optimal hybrid analog and
  digital precoding for downlink mmwave massive {MIMO} systems,'' in
  \emph{Proc. IEEE ICC 2015}, London, UK, Jun. 2015, pp. 1334--1339.

\bibitem{Li2017}
A.~Li and C.~Masouros, ``Hybrid analog-digital millimeter-wave mu-mimo
  transmission with virtual path selection,'' \emph{IEEE Commun. Lett.},
  vol.~21, no.~2, pp. 438--441, Feb. 2017.

\bibitem{Molu2018}
M.~M. Molu, P.~Xiao, M.~Khalily, K.~Cumanan, L.~Zhang, and R.~Tafazolli,
  ``Low-complexity and robust hybrid beamforming design for multi-antenna
  communication systems,'' \emph{IEEE Trans. Wireless Commun.}, vol.~17, no.~3,
  pp. 1445--1459, Mar. 2018.

\bibitem{Yu2016}
X.~Yu, J.-C. Shen, J.~Zhang, and K.~B. Letaief, ``Alternating minimization
  algorithms for hybrid precoding in millimeter wave {MIMO} systems,''
  \emph{IEEE J. Sel. Topics Signal Process.}, vol.~10, no.~3, pp. 485--500,
  Feb. 2016.

\bibitem{He2018}
L.~He, J.~Wang, and J.~Song, ``Spatial modulation for more spatial
  multiplexing: {RF}-chain-limited generalized spatial modulation aided mm-wave
  {MIMO} with hybrid precoding,'' \emph{{IEEE} Trans. Commun.}, vol.~66, no.~3,
  pp. 986--998, Mar. 2018.

\bibitem{Mesleh2008}
R.~Mesleh, H.~Haas, S.~Sinanovic, C.~W. Ahn, and S.~Yun, ``Spatial
  modulation,'' \emph{{IEEE} Trans. Veh. Technol.}, vol.~57, no.~4, pp.
  2228--2241, Jul. 2008.

\bibitem{Yang2008}
Y.~Yang and B.~Jiao, ``Information-guided channel-hopping for high data rate
  wireless communication,'' \emph{{IEEE} Commun. Lett.}, vol.~12, no.~4, pp.
  225--227, Apr. 2008.

\bibitem{Renzo2014}
M.~D. Renzo, H.~Haas, A.~Ghrayeb, S.~Sugiura, and L.~Hanzo, ``Spatial
  modulation for generalized {MIMO}: Challenges, opportunities, and
  implementation,'' \emph{Proc. {IEEE}}, vol. 102, no.~1, pp. 56--103, Jan.
  2014.

\bibitem{Guo2016}
S.~Guo, H.~Zhang, S.~Jin, and P.~Zhang, ``Spatial modulation via {3-D}
  mapping,'' \emph{{IEEE} Commun. Lett.}, vol.~20, no.~6, pp. 1096--1099, Jun.
  2016.

\bibitem{Guo2017}
S.~Guo, H.~Zhang, P.~Zhang, D.~Wu, and D.~Yuan, ``Generalized {3-D}
  constellation design for spatial modulation,'' \emph{{IEEE} Trans. Commun.},
  vol.~65, no.~8, pp. 3316--3327, Aug. 2017.

\bibitem{Ibrahim2016}
A.~A.~I. Ibrahim, T.~Kim, and D.~J. Love, ``On the achievable rate of
  generalized spatial modulation using multiplexing under a {Gaussian} mixture
  model,'' \emph{{IEEE} Trans. Commun.}, vol.~64, no.~4, pp. 1588--1599, Apr.
  2016.

\bibitem{Basnayaka2016}
D.~A. Basnayaka, M.~D. Renzo, and H.~Haas, ``Massive but few active {MIMO},''
  \emph{{IEEE} Trans. Veh. Technol.}, vol.~65, no.~9, pp. 6861--6877, Sep.
  2016.

\bibitem{Faregh2016}
E.~Faregh, M.~Eslami, and J.~Haghighat, ``A closed-form mutual information
  approximation for multiple-antenna systems with spatial modulation,''
  \emph{IEEE Wireless Communications Letters}, vol.~5, no.~6, pp. 636--639,
  Dec. 2016.

\bibitem{Narasimhan2016}
T.~L. Narasimhan and A.~Chockalingam, ``On the capacity and performance of
  generalized spatial modulation,'' \emph{{IEEE} Commun. Lett.}, vol.~20,
  no.~2, pp. 252--255, Feb. 2016.

\bibitem{Younis2017}
A.~Younis, N.~Abuzgaia, R.~Mesleh, and H.~Haas, ``Quadrature spatial modulation
  for {5G} outdoor millimeter�cwave communications: Capacity analysis,''
  \emph{IEEE Trans. Wireless Commun.}, vol.~16, no.~5, pp. 2882--2890, May
  2017.

\bibitem{Younis2018}
A.~Younis and R.~Mesleh, ``Information-theoretic treatment of space modulation
  {MIMO} systems,'' \emph{{IEEE} Trans. Veh. Technol.}, vol.~67, no.~8, pp.
  6960--6969, Aug. 2018.

\bibitem{Guo2019}
\BIBentryALTinterwordspacing
S.~Guo, H.~Zhang, P.~Zhang, P.~Zhao, L.~Wang, and M.-S. Alouini, ``Generalized
  beamspace modulation using multiplexing: A breakthrough in mm-wave {MIMO},''
  \emph{submitted to IEEE J. Sel. Areas Commun.}, Nov. 2018. [Online].
  Available: \url{https://arxiv.org/pdf/1810.10724.pdf}
\BIBentrySTDinterwordspacing

\bibitem{Liu2016}
C.~Liu, M.~Ma, Y.~Yang, and B.~Jiao, ``Optimal spatial-domain design for
  spatial modulation capacity maximization,'' \emph{{IEEE} Commun. Lett.},
  vol.~20, no.~6, pp. 1092--1095, Jun. 2016.

\bibitem{Cover2012}
T.~M. Cover and J.~A. Thomas, \emph{Elements of information theory}.\hskip 1em
  plus 0.5em minus 0.4em\relax John Wiley \& Sons, 2012.

\bibitem{Huber2008}
M.~F. Huber, T.~Bailey, H.~Durrant-Whyte, and U.~D. Hanebeck, ``On entropy
  approximation for {Gaussian} mixture random vectors,'' in \emph{Proc. IEEE
  Int. Conf. Multisensor Fusion Integr. Intell. Syst. (MFI),}.\hskip 1em plus
  0.5em minus 0.4em\relax Seoul, Korea: IEEE, Aug. 2008, pp. 181--188.

\bibitem{Shi2017}
Y.~Shi, M.~Ma, Y.~Yang, and B.~Jiao, ``Optimal power allocation in spatial
  modulation systems,'' \emph{{IEEE} Trans. Wireless Commun.}, vol.~16, no.~3,
  pp. 1646--1655, Mar. 2017.

\end{thebibliography}
\end{document}